# Disorder-induced linear magnetoresistance in Al$_2$O$_3$/SrTiO$_3$ heterostructures

Kuang Hong Gao,[1,*] Tie Lin,[2,†] Xiao Rong Ma,[1] Qiu Lin Li[1], and Zhi Qing Li[1]

[1]*Tianjin Key Laboratory of Low Dimensional Materials Physics and Preparing Technology, Department of Physics, Tianjin University, Tianjin 300354, China*

[2]*State Key Laboratory of Infrared Physics, Chinese Academy of Sciences, Shanghai Institute of Technical Physics, Shanghai 200083, China*

**Abstract**

An unsaturated linear magnetoresistance (LMR) has attracted widely attention because of potential applications and fundamental interest. By controlling growth temperature, we realized a metal-to-insulator transition in Al$_2$O$_3$/SrTiO$_3$ heterostructures. The LMR is observed in metallic samples with electron mobility varying over three orders of magnitude. The observed LMR cannot be explained by the guiding center diffusion model even in samples with very high mobility. The slope of the observed LMR is proportional to Hall mobility, and the crossover field, indicating a transition from quadratic (at low fields) to linear (at high fields) field dependence, is proportional to the inverse Hall mobility. This signifies that the classical model is valid to explain the observed LMR. More importantly, we develop an analytical expression according to the effective-medium theory that is equivalent to the classical model. And the analytical expression describes the LMR data very well, confirming the validity of the classical model.

---

[*] Corresponding author, e-mail: khgao@tju.edu.cn
[†] Corresponding author, e-mail: lin_tie@mail.sitp.ac.cn

# I. INTRODUCTION

A linear increase of resistance with external magnetic field is defined as linear magnetoresistance (LMR). It has attracted much attention for technological applications in magnetic sensors and memory devices as well as for fundamental development to probe the band topology. The LMR has been observed in a wide class of materials including metal, semimetal, and superconductors [1-2]. It is unsaturated in strong magnetic fields, which is different from conventional magnetoresistance that is quadratic in a low field range and quickly saturates to a small value at high fields [3]. Theoretically, Abrikosov proposed a quantum model to explain the unsaturating LMR, in which it should be observed when only the lowest Landau level is occupied by carriers in Dirac materials with zero bandgap and linear dispersion [4]. Apparently, the realization of this quantum LMR requires materials with low carrier concentration and/or the exertion of high external magnetic field. This mechanism is reported to be responsible for the appearance of LMR in multilayer graphene [5], iron pnictides [6], bismuth chalcogenides [7], ZrGeSe [8], and bismuth [9]. Alternatively, Parish et al. proposed a classical model that releases the above restriction condition [10]. In this classical model, the LMR originates from disorder: the strong disorder is predicted to be critical to realize significant LMR. This model has been widely used to explain the observation of LMR in narrow and zero bandgap materials such as InSb [11], monolayer graphene [12], Dirac semimetals [13], and topological insulators [14-15]. In conducting materials with very high mobility where disorder is expected to be suppressed, however, this mechanism is challenged by a semi-classical guiding center diffusion model [16] in which the LMR results from long mean free path and high carrier mobility (e.g., higher than 10,000 $cm^2V^{-1}s^{-1}$) [17]. Therefore, to explore the origin of LMR, it is necessary to find a material platform in which the LMR can be realized in a wide mobility range.

In this paper, $Al_2O_3/SrTiO_3$ heterostructures is found to be such a material. A series of high-quality $Al_2O_3/SrTiO_3$ heterostructures are prepared by magnetron sputtering. By controlling growth temperature, the LMR is observed in a wide mobility range of 444-130,841 $cm^2V^{-1}s^{-1}$. Experimentally, the LMR in $SrTiO_3$-based transport systems has been reported by several groups. Liu et al. firstly reported the LMR in $SrTiO_3$ single crystal reduced by annealing at high vacuum, and they attributed it to the quantum model [18]. In Nb-doped $SrTiO_3$ thin films, the appearance of LMR was explained by weak antilocalization [19]. In Ar-ion-irradiated $SrTiO_3$ single crystal [20], γ-$Al_2O_3$/$SrTiO_3$ and $GdO_x$/$SrTiO_3$ heterostructures [21-22], however, the large LMR was attributed to the classical model. These contradictive experimental results make it necessary to further investigate the LMR in $SrTiO_3$-based transport systems. Herein, we study the electrical transport properties of $Al_2O_3$/$SrTiO_3$ heterostructures. Because of close relation between the observed LMR and carrier's mobility, we found that the classical model can be applied to explain the LMR. Importantly, an analytical expression, obtained according to the effective-medium theory (EMT) [23], can be used to fit the observed LMR, which confirms the validity of the classical model.

## II. THEORY

In the classical model, the conducting system is phenomenologically regarded as a network of some small resistors. The resistance of each unit is random, which thus causes the spatial variation of conductivity. This variation of conductivity mimics the disorder effect that is described by fluctuating mobility while the inhomogeneity of carrier concentration is not considered. When external magnetic field $B$ is aligned perpendicular to the current, the unsaturating magnetoresistance shows a quadratic $B$ dependence at low fields and a linear $B$ dependence (i.e., the LMR) at high fields. The magnitude of LMR can be dominated either by mobility or by its fluctuation. According to the theory, two criteria are usually used to check its validity [24]: (1) the magnitude (equivalently, the slope) of LMR is proportional to either the Hall mobility or its fluctuation, and (2) the crossover field $B_C$, at which there occurs a transition from quadratic (at low fields) to linear (at high fields) $B$ dependence, is proportional to the inverse Hall mobility. Unfortunately, no analytical expression is given in the model for quantitatively analyzing the LMR.

For the disorder-induced LMR, another phenomenological model is the EMT. This theory assumes that a disordered system consists of many macroscopic puddles, each of which has a given conductivity. These puddles provide a varying disorder potential that leads to spatially varying local carrier concentrations. The EMT has been proven to be equivalent to the classical model and can be used to describe LMR in a wide class of materials [25]. Considering the fluctuation of carrier concentrations, effective conductivity tensor $\hat{\sigma}^E$ is given by following equation [25-26]

$$\int dn P(n) \frac{\hat{\sigma}(n,\mu,B) - \hat{\sigma}^E}{1 + (1/2\hat{\sigma}_{xx}^E)[\hat{\sigma}(n,\mu,B) - \hat{\sigma}^E]} = 0, \quad (1)$$

where $P(n)$ is the distribution function of carrier concentration $n$, $\hat{\sigma}$ is conductivity tensor, and $\mu$ is mobility. In general, $P(n)$ is taken to be a Gaussian distribution: $P(n) = \exp[(n-\bar{n})/\Delta n]^2$ (here, $\bar{n}$ is the mean carrier concentration and $\Delta n$ is the standard deviation). For this distribution, however, there are two disadvantages: (1) a case that $n < 0$ is inevitable, as seen in Fig. 1(a), which is physically meaningless, and (2) near $n = -\hat{\sigma}^E/(e\mu)$, we found that the left-hand side of Eq. (1) becomes divergent.

To address the problem, we adopt the elliptical distribution

$$P(n) = \frac{2}{\pi s \bar{n}} \sqrt{1 - \left(\frac{n-\bar{n}}{s\bar{n}}\right)^2}, \qquad |n-\bar{n}| < s\bar{n} \quad (2)$$

$$P(n) = 0. \qquad |n-\bar{n}| \geq s\bar{n} \quad (3)$$

Here, $s$ is a coefficient ($0 \leq s < 1$). When $s$ is small, it is possible to avoid that $n < 0$ [Fig. 1(a)]. Considering two relations $\hat{\sigma}(n,\mu,B) \equiv ne\mu/(1+i\mu B) \equiv \hat{\sigma}_{xx} - i\hat{\sigma}_{xy}$ and $\hat{\sigma}^E = \hat{\sigma}_{xx}^E - i\hat{\sigma}_{xy}^E$, we can obtain $\hat{\sigma}_{xx}^E$ and $\hat{\sigma}_{xy}^E$ according to Eq. (1) (see Appendix):

$$\hat{\sigma}_{xx}^E(B) = \frac{\bar{\sigma}_0}{1+\mu^2 B^2} \left[\frac{1}{2} + \frac{1}{2}\sqrt{1 - \frac{1}{2}s^2(1-\mu^2 B^2)}\right], \quad (4)$$

$$\hat{\sigma}_{xy}^E(B) = \frac{\bar{\sigma}_0 \mu B}{1+\mu^2 B^2} \left\{1 - \frac{s^2}{2[1+\sqrt{1-0.5s^2(1-\mu^2 B^2)}]}\right\}. \quad (5)$$

Here, $\bar{\sigma}_0 = \bar{n}e\mu$. Magnetoresistance (MR) can be defined as MR $= [\hat{\rho}_{xx}^E(B) - \hat{\rho}_{xx}^E(0)]/\hat{\rho}_{xx}^E(0)$ ($\hat{\rho}_{xx}^E$ is the effective resistivity tensor). Then, one can obtain

$$\text{MR} = \frac{\hat{\sigma}_{xx}^{\text{E}}(B) \cdot \hat{\sigma}_{xx}^{\text{E}}(0)}{\hat{\sigma}_{xx}^{\text{E}}(B)^2 + \hat{\sigma}_{xy}^{\text{E}}(B)^2} - 1, \qquad (6)$$

Equation (6), combined with Eqs. (4) and (5), can be used to fit experimental data with two fitting parameters of $\mu$ and $s$.

### III. EXPERIMENTAL METHOD

We deposited $Al_2O_3$ film on $SrTiO_3$ (100) substrate for forming heterostructures by magnetron sputtering. Prior to $Al_2O_3$ deposition, $SrTiO_3$ substrate has been etched in an acid solution and subsequently annealed in oxygen atmosphere to produce a $TiO_2$-terminated surface. After $Al_2O_3$ target (99.99%) was pre-sputtered for 20 minutes, sputtering was performed for 1 hour with a power of 80 W. During sputtering, the growth temperature in the range of 723-873 K was controlled by a heater installed on the back of sample holder. Five samples were referred to as samples S1-S5, the growth temperature for which increases in sequence (see Table 1). X-ray diffraction indicates that the grown films belong to $\alpha$-$Al_2O_3$ structure, which has been reported in a previous paper [27]. Hall-bar shaped samples were obtained by a shadow mask. For the measurement of electrical properties, aluminum was deposited by thermal evaporation to form Ohmic contacts. Magnetic-transport and Hall effect measurements were performed in a physical property measurement system (PPMS-6000, Quantum Design) with a magnetic field $B$ perpendicular to the plane of sample. We should emphasize that resistance in gap between Hall bar electrodes is highly insulating beyond measurement limitation (i.e., higher than 10 MΩ). This indicates that the occurrence of electrical conduction outside Hall bar is negligible. The Hall measurement suggests that all our samples are N-type, meaning that transport carriers are electrons.

### IV. RESULTS AND DISCUSSION

Figure 1(b) shows a cross-sectional transmission electron microscopy image for sample S2. An interface between $Al_2O_3$ and $SrTiO_3$ layers is visible, indicating that $Al_2O_3$/$SrTiO_3$ heterostructures are prepared by the magnetron sputtering. The thickness of the grown $Al_2O_3$ layer is 36 nm for sample S2 [see the inset of Fig. 1(b)], and similar values are also obtained for all other samples. If the polar discontinuity is responsible for the formation of two-dimensional electron gas near the interface, the $Al_2O_3$ layer with thickness larger than ~1.2 nm is indispensable for obtaining conducting interface [28]. Since the thickness of the grown $Al_2O_3$ layer is far larger than ~1.2 nm in our samples, a conducting interface in the $Al_2O_3/SrTiO_3$ heterostructures is expected. On the other hand, it has been experimentally reported that the formation of two-dimensional electron gas is likely to originate from oxygen vacancies near interface [29]. In this case, the critical thickness of ~1.2 nm does not exist, and oxygen vacancies are the source of transport electrons. This, together with the reported result that oxygen vacancies are prone to be introduced at high temperature [30], provides us an effective route to change the amount of oxygen vacancies and thus control electron concentration $n$ by simply changing the growth temperature. Indeed, as seen in Table 1, $n$ demonstrates a remarkable increase on increasing the growth temperature.

Figure 2(a) shows sheet resistance $R_S$ as a function of temperature $T$ for all our samples. One can see that $R_S$ increases with decreasing $T$ for sample S1, indicating an insulating behavior. On the contrary, $R_S$ decreases with decreasing $T$ from 300 K down to ~30 K for sample S2, that is, a positive $T$ coefficient is observed. Below ~30 K, there is a slight increase which may be related to electron-electron interaction and Kondo effect. For samples S3-S5, a positive $T$ coefficient is also observed in high $T$ range, but $R_S$ exhibits a saturating trend below ~10 K. The positive $T$ coefficient in high $T$ range for samples S2-S5 indicates that one can obtain metallic samples when the growth temperature is higher than 800 K. Considering the insulating behavior of sample S1, one can find that a metal-insulator transition is realized by controlling the growth temperature for our samples. According to the Ioffe−Regel criterion [31], a disordered system should exhibit metallic (insulating) character when $k_F l > 1$ ($k_F l < 1$) (here, $k_F$ is the Fermi wave number and $l$ is the mean free path). That is, when $k_F l$ is close to unity, a metal-insulator transition is expected. Since $k_F l \approx (25\,[k\Omega])/R_S$ in a two-dimensional system, that $R_S = 25\,k\Omega$ becomes a critical value for the metal-insulator transition [32]. We note that the $R_S$ of sample S1 is larger than 25 kΩ while that is smaller than 25 kΩ for other samples over the studied $T$ range. This explains why sample S1 exhibits the insulating character while samples S2-S5 are metallic. For the metallic samples, residual resistivity ratio (RRR) varies between 147 and 5003. The RRR value reaching to 5003 is indicative of high sample quality. Below, we mainly discuss the electrical transport properties of metallic samples.

Figure 2(b) shows magnetoresistance (MR) at 2 K for samples S2-S5. Near zero fields ($|B|<1.0$ T), weak antilocalization-induced dip appears in sample S2 and only positive parabolic MR is observed in samples S3-S5. In high $B$ range, a well-defined LMR is observed in all metallic samples. But no quantum MR oscillations are discernible. This may be due to high measured $T$ (2-300 K) and low quantum mobility [33-36]. Importantly, it is noteworthy that the LMR is gradually enhanced for samples S2-S5 in sequence, and a maximum of 744% is obtained in sample S5 under $B = 9$ T. Since there is no linear dispersion in $Al_2O_3/SrTiO_3$ heterostructures [28,37] and very high electron concentration is obtained especially in samples S3 and S4, the observed LMR cannot be attributed to the quantum model. To clarify mechanism responsible for the LMR, it is necessary to obtain the Hall mobility. We note that Hall resistance is nonlinear when $T < 40$ K, which may result from a double-subband occupation of the different types of d$xy$ and d$xz/yz$ subbands [38]. Then, total electron concentration $n$ is determined from high-field Hall coefficient [21]. From the determined $n$, effective Hall mobility $\mu_H$ is calculated using a relation $1/R_{S0} = ne\mu_H$ ($R_{S0}$ is sheet resistance under $B = 0$ T). As seen in Table 1, the calculated $\mu_H$ at 2 K rapidly increases for samples S2-S5 in sequence. Apparently, there is a positive correlation between the magnitude of LMR and $\mu_H$ value. This correlation is predicted both by the classical model and by the guiding center diffusion model. But we find that the latter cannot explain the observed LMR in our samples for two reasons. First, the guiding center diffusion model requires that transport time $\tau_{tr}$ is much greater than cyclotron period $2\pi/\omega_c$ (i.e., $\tau_{tr} \gg 2\pi/\omega_c$ )[16]. Considering that $\tau_{tr} = \mu_H m^*/e$ and $\omega_c = m^*/(eB)$ ($m^*$ is the effective mass of electron), one can find that the condition $\tau_{tr} \gg 2\pi/\omega_c$ is equivalent to an inequality $\mu_H \gg 2\pi/B$. That is, for the highest field of $B = 9$ T, the guiding center diffusion model applies only when $\mu_H \gg 698$

cm$^2$V$^{-1}$s$^{-1}$. Naturally, the LMR should not appear in sample S2 with $\mu_H < 698$ cm$^2$V$^{-1}$s$^{-1}$. This is inconsistent with observation in Fig. 2(b). Furthermore, the LMR has been well developed at $B = 2$ T for samples S3 and S4. But the condition that $\mu_H \gg 2\pi/B$ with $B = 2$ T (i.e., $\mu_H \gg 31,416$ cm$^2$V$^{-1}$s$^{-1}$) is not satisfied in these two samples (see Table 1). This further supports that the guiding center diffusion model cannot explain the observed LMR in our samples. Second, this model predicts that the $B$ dependence of resistivity can be calculated by $\rho_{xx}(B) = (\mu_H[\text{cm}^2\text{V}^{-1}\text{s}^{-1}]/10^4) \cdot B[\text{T}] \cdot \tan\theta_H / (1 + \tan^2\theta_H)$ where $\tan\theta_H$ is Hall angle. However, we find that the calculated MR is not consistent with experimental data. Taking sample S4 as an example, the calculated MR under $B = 9$ T is equal to ~30,000% that is far larger than the experimental value of 544%. It can be concluded that the observed LMR cannot be attributed to the guiding center diffusion model although very high mobility is realized in our samples.

This leaves the classical model as the most likely mechanism responsible for the LMR. Before this model is used to analyze the experimental data, one should note that it sets an upper limit on changes in longitudinal resistance $R_{xx}$: $\Delta R_{xx} < |R_{xy}|$ (here, $\Delta R_{xx} = R_{xx}(B) - R_{xx}(0)$ and $R_{xy}$ is Hall resistance) [8]. Indeed, the condition $\Delta R_{xx} < |R_{xy}|$ is found to be satisfied in our samples, although the high LMR reaching to 744% is obtained. Figure 3(a) shows $-R_{xy}$ and $\Delta R_{xx}$ as a function of $B$ for sample S5 as an example (here, $-R_{xy} = |R_{xy}|$ due to the N-type of our samples). One can see that $\Delta R_{xx}$ value is less than $-R_{xy}$ in high $B$ range, meaning that the upper limit is not broken in our samples. Figure 3(b) shows the first derivative $d(\text{MR})/dB$ of LMR near $B = 9$ T as a function of $\mu_H$ at 2 K for samples S2-S5. One can see that $d(\text{MR})/dB$ is proportional to $\mu_H$, consistent with the classical model. Figure 3(c) shows the $T$ dependence of MR data for sample S5. Well-defined LMR is observed in high $B$ range for every fixed $T$, but it is gradually suppressed on increasing $T$. Similar phenomena are also observed in other metallic samples except for sample S2 at low $T$s presumably due to the presence of weak antilocalization. Figure 3(d) shows $d(\text{MR})/dB$ as function of $\mu_H$ for sample S5. One can see that the $d(\text{MR})/dB$ scales linearly with $\mu_H$. This, together with data in Fig. 3(b), strongly indicates that $d(MR)/dB \propto \mu_H$, meaning that the observed LMR in our samples has a classical origin.

This can be further confirmed by the other aforementioned criterion about relation between $B_C$ and $\mu_H^{-1}$. Figure 4(a) shows the $d(MR)/dB$ as a function of $B$ for samples S2-S5 at 2 K, where $B_C$ is marked by filled circles. The $B_C$, as seen in Fig. 4(b), is proportional to the $\mu_H^{-1}$. This is indicative of the applicability of the classical model in our samples. On the basis of the classical model, Kozlova et al. [39] considered the microscopic nature of the electron dynamics and found that the LMR arises from the stochastic movement of high-mobility electrons with cycloidal trajectories around low-mobility islands. These low-mobility islands are the source of disorder in transport system. In this picture,

$$d(MR)/dB = \mu_H f / 2(1-f), \qquad (7)$$

where $f$ is the fraction of the low-mobility islands. According to this expression, one can find that both strong disorder (correspondingly, high fraction of the low-mobility islands) and high carrier mobility are indispensable for the enhancement of classical LMR, although it seems to be difficult to simultaneously satisfy these two conditions. Substituting $d(MR)/dB$ and $\mu_H$ at 2 K into Eq. (7), one can obtain $f$ value. As shown in Table 1, the obtained $f$ values for our samples are comparable to the reported results of 32% for In(AsN) film [36], 40% for $GdO_x/SrTiO_3$ heterostructures [22], and 45%-70% for monolayer graphene [40]. Importantly, the obtained $f$ value decreases for samples S2-S5 in sequence. This suggests that the disorder becomes gradually suppressed for samples S2-S5 in sequence.

This point can be confirmed by applying the EMT theory to analyze the MR data. We used Eq. (6) combined with Eqs. (4) and (5) to fit the experimental data with two fitting parameters of $\mu$ and $s$. Here, it should be mentioned that parameter $\mu$ corresponds to the effective mobility because a single subband is considered in the EMT theory. The similar fitting procedure is also reported in Ref. [26]. As shown in Fig. 2(b) for the MR of samples S2-S5 at 2 K, the fitting curves (red solid lines) follow the experimental data very closely. Similarly, the $T$ dependence of the MR data for sample S5, as seen in Fig. 3(c), is also well described by the EMT theory. The fit-obtained $\mu$, as shown in Fig. 4(c), is consistent with $\mu_H$ for all metallic samples, indicating the validity of our fits. Figure 4(d) shows the extracted $s$ as a function of $T$ for sample S2-S5. All the extracted $s$ values vary between 0.15 and 0.75 (i.e., $0 < s < 1$), which self-consistently suggests the validity of our fits again. This confirms that the classical model can be applied to our samples because of the equivalence of this model and the EMT. For sample S5, the extracted $s$ (closed triangles) is $T$-independent below 30 K. But it exhibits an increasing trend with elevating $T$ from 30 K up to 150 K, above which MR is too weak to be reliably fitted. The reason for increase in $s$ at high $T$s will be discussed later. For other samples, the extracted $s$ is also $T$-independent below 30 K. More importantly, it exhibits a decreasing trend for samples S2-S5 in sequence below 30 K. This signifies that disorder is indeed gradually suppressed, consistent with the decreases in $f$.

For the suppression of disorder, there are two possible reasons. First, note that the disorder can be partially screened by high electron concentration. Since the electron concentration remarkably increases for samples S2-S5 in sequence (see Table 1), the disorder is expected to be suppressed by gradually enhanced screening. Second, the high electron concentration especially in samples S4 and S5 enables some electrons to escape from the interface and enter into $SrTiO_3$ substrate. As a consequence, a new transport channel is formed within the $SrTiO_3$ substrate. It is reasonable to speculate that the newly formed transport channel must be cleaner than that confined in interface, which is thus effectively suppress the disorder for overall electrical transport behavior.

In addition, the aforementioned increase of the extracted $s$ in the $T$ range of 30-150 K for sample S5 [see Fig. 4(d)] implies that disorder is enhanced at high $T$s. But we cannot obtain the $s$ value at high $T$s in other samples because MR becomes very weak above 50 K. To confirm the enhancement of disorder, we calculated $f$ value in sample S5. The calculated $f$ is found to increase at high $T$s (not given here), proving the enhanced disorder. For the enhanced disorder in the $T$ range of 30-150 K, there are also two possible reasons. First, we note that the dielectric constant of SrTiO$_3$ has been reported to be $T$-independent below ~50 K, above which it rapidly decreases [41]. The decrease in dielectric constant above ~50 K is expected to weaken the screening for electrons. As a result, the disorder is enhanced to some extent. Second, electron-electron scattering has been reported to become important in the $T$ range of 30-150 K [42]. Hence, this scattering may also have non-negligible contribution to the enhancement of disorder.

What is the source of disorder? Oxygen vacancy is believed to be most possible candidate. In lightly doped SrTiO$_3$, Collignon et al. claimed that the oxygen vacancy-induced mesoscopic dipoles scatter carriers and thus give rise to quasilinear MR [43]. However, notice that the quasilinear MR is diminished on increasing carrier concentration, which is contrary to our observations in Fig. 2(b). Therefore, the observed LMR in our samples cannot be explained by the formation of mesoscopic dipoles. On the other hand, the oxygen vacancy distribution has been experimentally found to be not uniform in SrTiO$_3$ [44], and some neighboring oxygen vacancies can form cluster [45]. Considering that oxygen vacancy is prone to be located near the interface of Al$_2$O$_3$/SrTiO$_3$ heterostructures [28], one can speculate that the inhomogeneity of oxygen vacancies near the interface must be the main source of disorder. Increasing $n$ for samples S2-S5 in sequence is suggestive of increase in oxygen vacancy concentration that may enhance disorder. The enhancement of disorder will compete with the screening arising from the high electron concentration. The screening must overwhelm the former, which thus induce the aforementioned suppression of disorder for samples S2-S5 in sequence.

It should be mentioned that the high electron concentration especially in samples S4 and S5 makes some electrons enter into SrTiO$_3$ substrate, and consequently a three dimensional transport is expected. It is worth noting that Eqs. (4)-(6) are valid only in two dimensional case. The observation that the LMR can be well described by Eqs. (4)-(6) in samples S4 and S5 indicates that two-dimensional electron gas confined in the interface dominates the magneto-transport behavior. On the other hand, as discussed before, the classical model uses the fluctuating mobility to scale the disorder. In contrast, Eqs. (4)-(6) from the EMT consider the inhomogeneity of carrier concentration to describe the LMR very well. This indicates that the inhomogeneity of carrier concentration in the EMT plays the same role in scaling the disorder as the fluctuating mobility in the classical model.

## V. CONCLUSION

We study the electrical transport properties of $Al_2O_3$/$SrTiO_3$ heterostructures grown by magnetron sputtering. The LMR is observed in samples with mobility varying over three orders of magnitude. The classical model is found to be valid to explain the observed LMR even with mobility reaching to 130,841 $cm^2V^{-1}s^{-1}$, while the guiding center diffusion model cannot be applied to our samples. The validity of the classical model is further confirmed by the observation that the LMR can be well described by the analytical expression obtained from the EMT.

## ACKNOWLEDGMENTS

This work was supported by National Natural Science Foundation of China (Grant Nos. 12174282 and 61974153).

## APPENDIX: THE DERIVATION OF ANALYTICAL FORMULA

In this Appendix we show the derivation process of Eqs. (4) and (5). According to Eq. (1) of the effective-medium theory (EMT), we can obtain

$$\left\langle \frac{\hat{\sigma}(n,\mu,B)-\hat{\sigma}^E}{1+(1/2\hat{\sigma}^E_{xx})[\hat{\sigma}(n,\mu,B)-\hat{\sigma}^E]} \right\rangle = 0, \tag{A1}$$

where $\hat{\sigma}(n,\mu,B) \equiv \hat{\sigma}_{xx} - i\hat{\sigma}_{xy}$ and $\hat{\sigma}^E = \hat{\sigma}^E_{xx} - i\hat{\sigma}^E_{xy}$. This equation can be simplified to

$$2\hat{\sigma}^E_{xx} \left[ 1 - 2\hat{\sigma}^E_{xx} \left\langle \frac{1}{\hat{\sigma}^{E*} + \hat{\sigma}(n,\mu,B)} \right\rangle \right] = 0, \tag{A2}$$

where $\hat{\sigma}^{E*}$ is an adjoint of $\hat{\sigma}^E$. Because $\hat{\sigma}^E_{xx} \neq 0$, one can find

$$\left\langle \frac{1}{\hat{\sigma}^{E*} + \hat{\sigma}(n,\mu,B)} \right\rangle = \frac{1}{2\hat{\sigma}^E_{xx}}. \tag{A3}$$

For the elliptical distribution, we define $x \equiv (n-\bar{n})/s\bar{n}$, $\bar{\sigma} \equiv \bar{n}e\mu/(1+i\mu B)$, and $F \equiv \bar{\sigma}/(\hat{\sigma}^{E*}+\bar{\sigma})$. Then, Eq. (A3) is translated into

$$\frac{1}{\hat{\sigma}^{E*}+\bar{\sigma}}\left\langle\frac{1}{1+F\cdot s\cdot x}\right\rangle = \frac{1}{2\hat{\sigma}_{xx}^{E}}. \tag{A4}$$

Considering that $P(n)$ is an even function, we find that the term

$$\left\langle\frac{1}{1+F\cdot s\cdot x}\right\rangle = \frac{2}{\pi}\int_{-1}^{1}\frac{\sqrt{1-x^2}}{1+F\cdot s\cdot x}dx = \frac{2}{1+\sqrt{1-(F\cdot s)^2}}. \tag{A5}$$

Substituting this term into Eq. (A4), one can obtain

$$8\hat{\sigma}_{xx}^{E}(\hat{\sigma}^{E}-\bar{\sigma})+\bar{\sigma}^2 s^2 = 0. \tag{A6}$$

Due to real number $\bar{\sigma}_0 = \bar{n}e\mu$, Eq. (A6) is rewritten as

$$8\hat{\sigma}_{xx}^{E}(\hat{\sigma}_{xx}^{E}-i\hat{\sigma}_{xy}^{E}-\frac{\bar{\sigma}_0}{1+i\mu B})+s^2(\frac{\bar{\sigma}_0}{1+i\mu B})^2 = 0. \tag{A7}$$

Let both real and imaginary parts of the left-hand side of Eq. (A7) be equal to zero and one can obtain Eqs. (4) and (5), respectively.

(2012).

[40] K. H. Gao, W. J. Wang, T. Lin, and Z. Q. Li, Disorder-controlled linear magnetoresistance in monolayer graphene, Europhys. Lett. **142**, 26002 (2023).

[41] J. H. Hao, Z. Luo, and J. Gao, Effects of substrate on the dielectric and tunable properties of epitaxial SrTiO$_3$ thin films, J. Appl. Phys. **100**, 114107 (2006).

[42] D. V. Christensen, Y. Frenkel, P. Schutz, F. Trier, S. Wissberg, R. Claessen, B. Kalisky, A. Smith, Y. Z. Chen, and N. Pryds, Electron mobility in $\gamma$-Al$_2$O$_3$/SrTiO$_3$ interface, Phys. Rev. Appl. **9**, 054004 (2018).

[43] C. Collignon, Y. Awashima, Ravi, X. Lin, C. W. Rischau, A. Acheche, B. Vignolle, C. Proust, Y. Fuseya, K. Behnia, and B. Fauque, Quasi-isotropic orbital magnetoresistance in lightly doped SrTiO$_3$, Phys. Rev. Mater. **5**, 065002 (2021).

[44] Z. Q. Liu, D. P. Leusink, X. Wang, W. M. Lu, K. Gopinadhan, A. Annadi, Y. L. Zhao, X. H. Huang, S. W. Zeng, Z. Huang, A. Srivastava, S. Dhar, T. Venkatesan, and Ariando, Metal-insulator transition in SrTiO$_{3-x}$ thin films induced by frozen-out carriers, Phys. Rev. Lett. **107**, 146802 (2011).

[45] K. Eom, E. Choi, M. Choi, S. Han, H. Zhou, and J. Lee, Oxygen vacancy linear clustering in a perovskite oxide, J. Phys. Chem. Lett. **8**, 3500 (2017).
18

**Figure captions**

**Figure 1** (a) Gaussian distribution (solid line) and elliptical distribution (dashed line) function $P(n)$ versus carrier concentration $n$. (b) The cross-sectional TEM image of sample S2.

**Figure 2** (a) Sheet resistance $R_S$ as a function of temperature $T$ at magnetic field $B = 0$ T for all our samples. The dashed line corresponds to 25 kΩ. (b) Magnetoresistance (MR) at 2 K for samples S2-S5. Red solid lines are fits according to the EMT theory [23].

**Figure 3** (a) Hall resistance $R_{xy}$ and longitudinal resistance $\Delta R_{xx} = R_{xx}(B) - R_{xx}(0)$ as a function of $B$ for sample S5. (b) First derivative of linear MR, i.e., $d(MR)/dB$, in high field range as a function of Hall mobility $\mu_H$ at 2 K for samples S2-S5. Solid line provides a guide to the eye. (c) MR at various $T$s for sample S5. Red solid lines are fits according to the EMT theory [23]. (d) $d(MR)/dB$ as a function of $\mu_H$ for different $T$ for sample S5. Solid line is a guide to the eye.

**Figure 4** (a) $d(MR)/dB$ as a function of $B$ for different samples at 2 K. The crossover field $B_C$ is marked by filled circles. (b) $B_C$ as a function of the inverse of $\mu_H$ at 2 K for samples S2-S5. Solid line provides a guide to the eye. (c) A comparison between $\mu_H$ (dashed lines) and the extracted values (symbols) for samples S2-S5 in the $T$ range of 2-300 K. (d) The extracted $s$ as a function of $T$ for samples S2-S5 according to Eq. (6). Dashed lines provide guide to the eye for the data below 30 K.

**Table 1:** Growth condition and electrical parameters. $T_G$ is growth temperature. $n$ is electron concentration at 300 K. $u_H$ is Hall mobility at 2 K. $f$ is the fraction of the low-mobility islands in sample determined from Eq. (7) according to Ref. [36].

| Sample | $T_G$ (K) | $n$ ($10^{12}$ cm$^{-2}$) | $u_H$ (cm$^2$V$^{-1}$s$^{-1}$) | $f$ (%) |
|---|---|---|---|---|
| S1 | 723 | 8.43 | - | - |
| S2 | 823 | 257 | 444 | 70.5 |
| S3 | 833 | 1,650 | 16,986 | 7.9 |
| S4 | 853 | 2,700 | 50,056 | 20.4 |
| S5 | 873 | 25,100 | 130,841 | 11.6 |

Figure 1

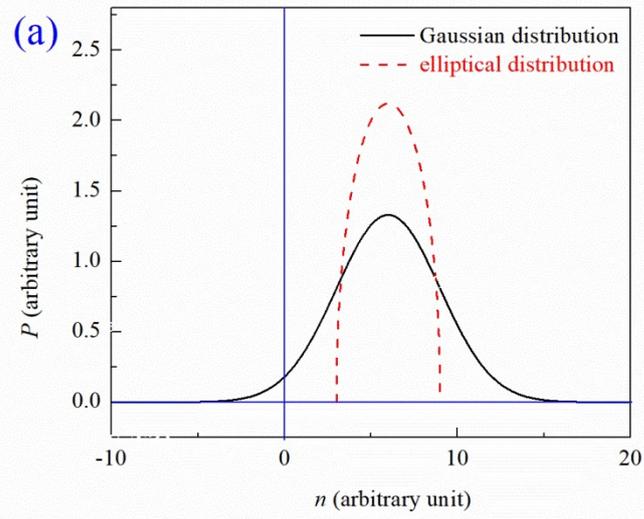
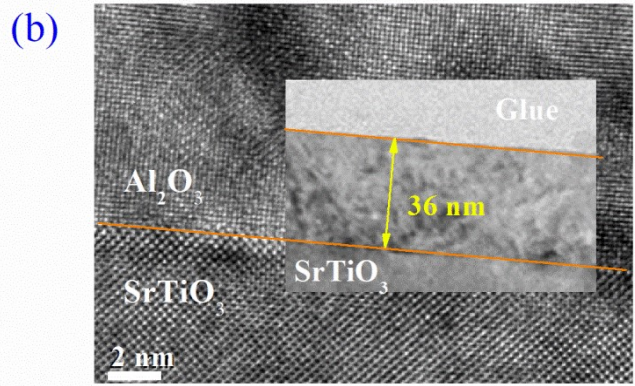
21

Figure 2

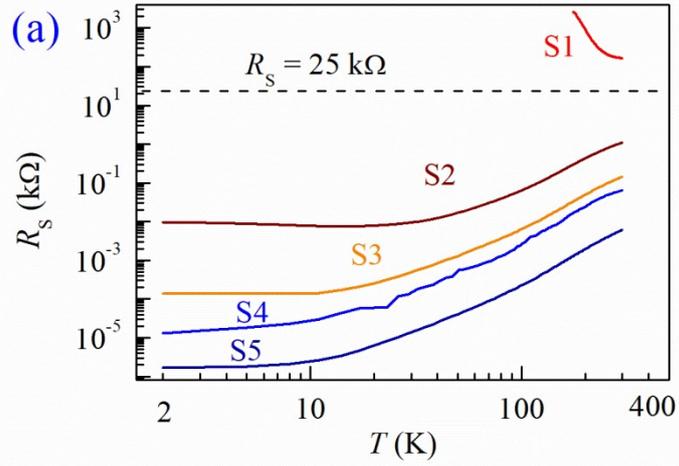

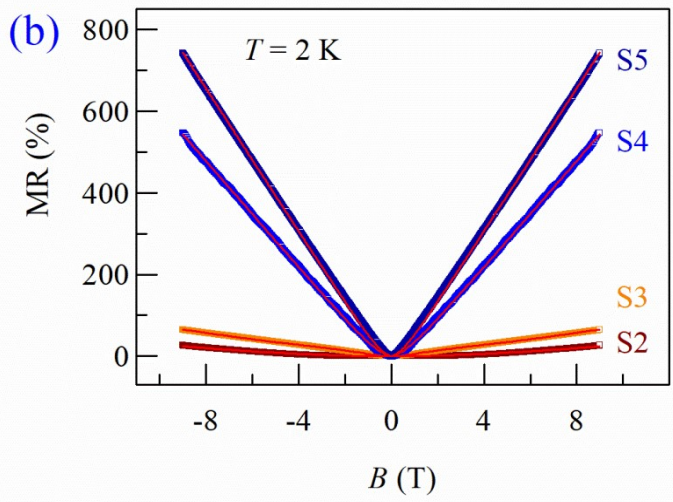

Figure 3

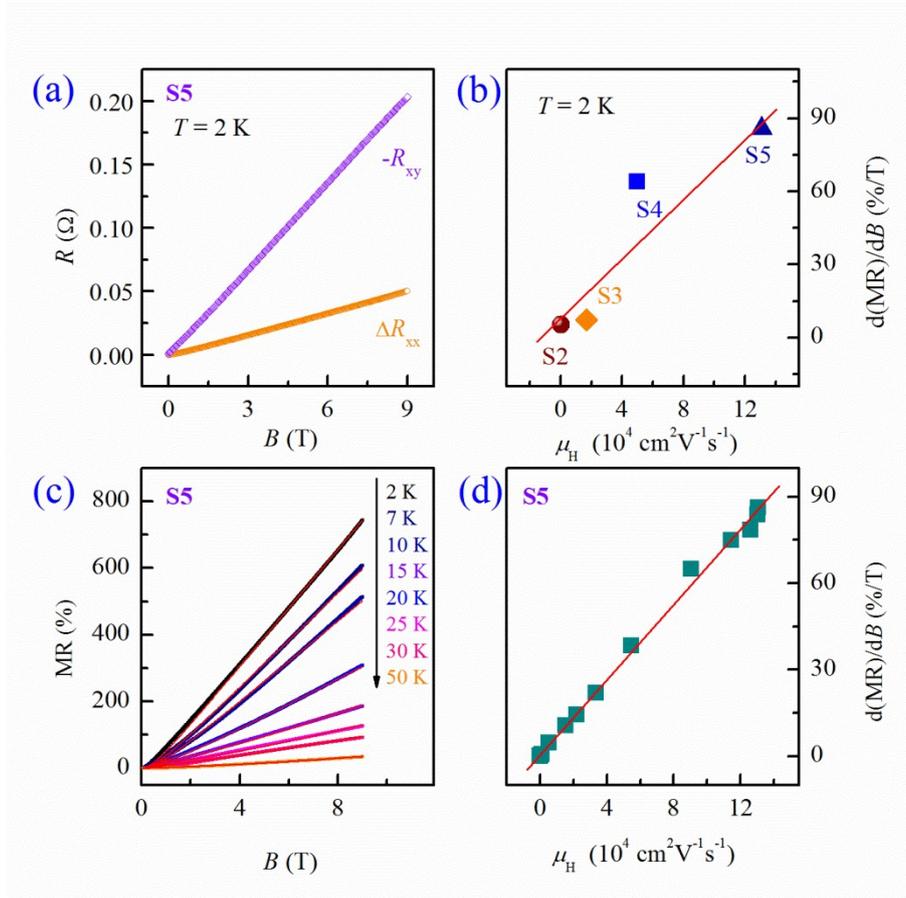

Figure 4

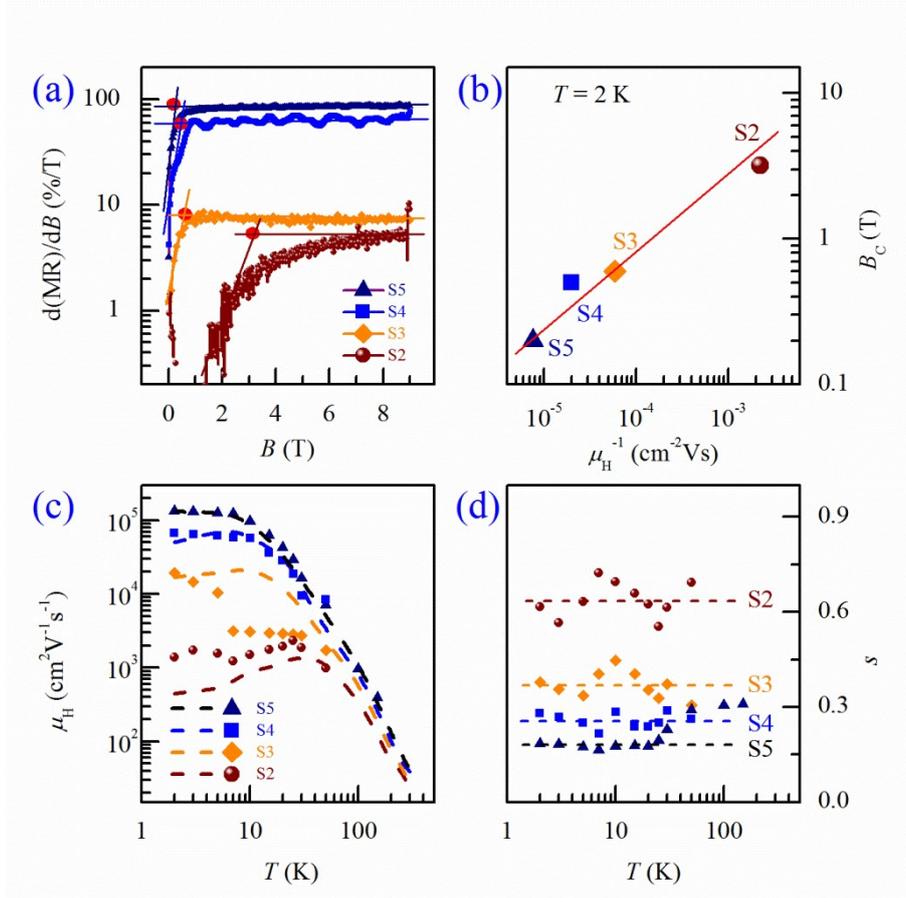